\documentclass[prb,preprint]{revtex4-1} 

\usepackage{color}
\usepackage{amsmath}  
\usepackage{amsfonts} 
\usepackage{graphicx} 
\usepackage{epstopdf} 

\begin{document}

\title{On the physics of propagating Bessel modes in cylindrical waveguides\\ \textcolor{blue}{Publication in process in American Journal of Physics}}

\author{J. E. G\'{o}mez-Correa}
\email{jesus.gomezcrr@uanl.edu.mx} 
\altaffiliation[permanent address: ]{Avenida Universidad s/n. Ciudad Universitaria San Nicol\'{a}s de los Garza, Nuevo Le\'{o}n, M\'{e}xico} 
\affiliation{Facultad de Ingenier\'{i}a Mec\'{a}nica y El\'{e}ctrica, Universidad Aut\'{o}noma de Nuevo Le\'{o}n, C.P. 66451, Nuevo Le\'{o}n, M\'{e}xico}
\affiliation{Centro de Investigaci\'{o}n Cient\'{i}fica y de Educaci\'{o}n Superior de Ensenada, Unidad Monterrey, C.P. 66629, Nuevo Le\'{o}n, M\'{e}xico}

\author{S. E. Balderas-Mata}
\email{sandra.balderas@cucei.udg.mx}
\affiliation{Departamento de Electr\'onica, Universidad de Guadalajara, C.P. 44840, Guadalajara, Jalisco, M\'exico}

\author{V. Coello}
\email{vcoello@cicese.mx}
\affiliation{Centro de Investigaci\'{o}n Cient\'{i}fica y de Educaci\'{o}n Superior de Ensenada, Unidad Monterrey, C.P. 66629, Nuevo Le\'{o}n, M\'{e}xico}

\author{N. P. Puente}
\email{norma.puenterm@uanl.edu.mx}
\affiliation{Facultad de Ingenier\'{i}a Mec\'{a}nica y El\'{e}ctrica, Universidad Aut\'{o}noma de Nuevo Le\'{o}n, C.P. 66451, Nuevo Le\'{o}n, M\'{e}xico}

\author{J. Rogel-Salazar}
\email{j.rogel@physics.org}
\affiliation{Science and Technology Research Institute, School of Physics Astronomy and Mathematics, University of Hertfordshire, Hatfield, Herts., AL10 9AB, UK}
\affiliation{Blackett Laboratory, Department of Physics, Imperial College London, Prince Consort Road, London, SW7 2BZ, UK}

\author{S. Ch\'{a}vez-Cerda}
\email{sabino@inaoep.mx}
\affiliation{Departamento de \'{O}ptica, Instituto Nacional de Astrof\'{\i}sica, \'{O}ptica y Electr\'{o}nica, Apdo. Postal 51/216, Puebla, M\'{e}xico}
\affiliation{Centro de Investigaciones en \'Optica, C.P. 37150, Le\'{o}n, Gto., M\'{e}xico}


\date{\today}

\begin{abstract}
In this paper we demonstrate that using a mathematical physics approach (focusing the attention to the physics and using mathematics as a tool) it is possible to visualize the formation of the transverse modes inside a cylindrical waveguide. In opposition, the physical mathematics solutions (looking at the mathematical problem and then trying to impose a physical interpretation), when studying cylindrical waveguides yields to the Bessel differential equation and then it is argued that in the core are only the Bessel functions of the first kind those who describe the transverse modes. And the Neumann functions are deemed non physical due to its singularity at the origin and eliminated from the final description of the solution. In this paper we show, using a geometrical-wave optics approach, that the inclusion of this function is physically necessary to describe fully and properly the formation of the propagating transverse modes. Also, the field in the outside of a dielectric waveguide arises in a natural way.
\end{abstract}

\maketitle 

\section{Introduction}

When working on the model of a physical phenomenon, it is sometimes
easy to concentrate on understanding the intricacies of the
mathematical methods employed to obtain the solutions that attempt to
describe the problem at hand, distracting the attention on the 
significance of the actual physical process in question. Once we have 
been successful in getting the mathematical solution, the physical process 
itself may not be properly addressed and emphasized, and in the worst 
cases it may even get lost.

In this work we deal with the description of the modes in cylindrical waveguides, where typically the physical phenomena
involved are treated in such a way that the balance between the
mathematical methods used and the physical constraints of the problem 
is tipped towards the former, leading to a somewhat unsatisfactory 
physical description of the formation of the modes in the waveguides. 
In this regard, it is worth remembering the words of Sommerfeld in the 
preface of his book on Partial Differential Equations, which although 
dealt with some physics, its main subject was actually mathematics 
\cite{sommerfeld}: ``\emph{We do not really deal with mathematical 
physics, but physical mathematics; not with the mathematical 
formulation of physical facts, but with the physical motivation of 
mathematical methods. The oft-mentioned prestabilized harmony between 
what is mathematically interesting and what is physically important 
is met at each step and   lends an esthetic - I should like to say 
metaphysical - attraction to our subject}.'' Of course, the last 
statement can be interpreted in both ways.

In order to provide a framework for the work presented here, let us
exemplify the difference between the physical mathematics
(p-mathematics) and the mathematical physics (m-physics) approaches
alluded to by Sommerfeld and which we will use in the rest of this
paper.  Consider the problem of finding the transverse modes in a
planar dielectric waveguide, on the one hand, the p-mathematics
approach is used when presenting the mathematical solutions and
boundary conditions in different regions of the slab. These solutions
and boundary conditions are matched at each interface between regions
and the description is said to be obtained. On the other hand, the
m-physics approach will look instead at the physical wave as it
travels through each of the regions of the slab. We can then observe
how the wave is reflected and transmitted as it reaches an interface
between the media in question and then set the equations of the
mathematical model according to the boundary conditions. We would like
to note that both approaches attempt to describe the same phenomenon,
however they may emphasize different aspects of the description.

In the case of the example used above, the p-mathematics approach will
give the sine and cosine functions as solutions inside the waveguide,
and the arguments of these functions are set to satisfy the boundary
conditions. In a different manner, the m-physics approach provides a
more extensive physical picture, i.e. that within the waveguide there
will be traveling waves suffering reflections at both interfaces that
in turn will have transverse components with the same frequency but
traveling in opposite directions. In this way, for some particular
conditions, the transverse modes in waveguides happen to be transverse
standing waves also referred to as stationary waves \cite{Straton,
  marcuse2}.

The m-physics approach is sometimes discussed in the study of planar
waveguides, mostly for the cosine function solution \cite{GHATAK, Hayt, Corson} but, to the best of our knowledge, it has never been used for
cylindrical waveguides and optical fibers. For the latter two the
p-mathematics approach is the standard one presenting the solutions in
terms of the Bessel functions of the first and second kind, then
discarding the latter arguing that they are \emph{unphysical} or not the
proper ones since they are singular at the axis of the waveguide.
\cite{Lee, Marcuse, Inan, Ramo, Okamoto, Agrawal, iizuka, saleh, chen}
With such arguments, based on the mathematical properties of the
functions without further thinking on what these properties might
physically represent, unfortunately, the physics of what really might
be happening in the physical system of the cylindrical waveguides can
be lost.

In this work we endeavor to use the mathematical formulation of the
physical facts, i.e. the \emph{mathematical physics} approach, to
present a detailed physical analysis of the formation of modes in cylindrical waveguides. We
demonstrate physically, with the help of propagating wave analysis, that
the well-known Bessel modes are in fact the result of the interference
of the counter-propagating transverse components of traveling conical
waves. Contrary to the prevailing approach in the literature we show
that to fully describe the propagating nature of the wave field in the
core and the cladding, it is necessary to use both Hankel functions;
constructed by the complex superposition of the Bessel function of the
first kind $J_m$ and the Bessel function of the second kind, or
Neumann function, $N_m$. Moreover, the singular behavior of these
solutions at the origin, in particular of the Neumann functions, is
easily explained in clear physical terms. Finally, our mathematical
physics traveling-wave approach shows how the traveling wave described
by the Hankel function within the waveguide becomes, in a natural way,
an evanescent wave at the interface between the core and the cladding
of dielectric cylindrical waveguides as it occurs for the planar
waveguides.

\section{Mathematical physics of cylindrical waveguides}
\label{sec:MP_Cyl}
In the literature on electromagnetic theory studying wave propagation
in cylindrical waveguides one encounters that the transverse field is
described by the Bessel differential equation. The set of independent
solutions of this equation are the Bessel functions of the first kind,
$J_{m}$, as well as the Bessel functions of the second kind, also
known as Neumann functions, $N_{m}$.

Following the literature related to cylindrical waveguides, the
treatment usually follows what we are referring to as the
p-mathematics approach. In other words, inclined more towards the mathematical approach. It is commonly argued that a
general solution can be constructed by a linear combination of these
functions, namely $E\left(\rho\right)=AJ_{m}+BN_{m}$, where $A$ and
$B$ are constants \cite{Lee, Marcuse, Inan, Ramo, Okamoto, Agrawal,
  iizuka, saleh, chen}. This superposition is mathematically correct
but unfortunately it does not necessarily give a complete physical
insight to the problem at hand. The argument used is along the lines
of \textquotedblleft looking for the physical
solution\textquotedblright\ inside the cylindrical waveguide, which
leads to the conclusion that the constant $B$ in the superposition
above needs to be zero. This is because the Neumann function has to be
discarded due to the fact that, for any $m$, it has a singularity at
the origin $\rho=0$, diverging to minus infinity. This argument
considers that such behavior of the solution is inconsistent with
physical fields within the core of the waveguide and that, the only
``proper'' and physically allowed solution that is bounded is $J_{m}$,
i.e. the Bessel function of the first kind.

The main aim of this paper is to show that in a full
description and analysis of a cylindrical waveguide, the Neumann
functions become a natural part of the solution and furthermore,
their presence provide a whole physical picture of how the modes are formed in cylindrical waveguides. In order to show this, we first solve the wave equation or the
Helmholtz equation in cylindrical coordinates,
$\nabla^{2}E\left(\rho,\varphi,z\right)
+k^{2}E\left(\rho,\varphi,z\right) =0$, by applying separation of
variables and using
$E(\rho,\varphi,z)=R\left(\rho\right)\Phi\left(\varphi\right)Z\left(
  z\right) $ as an ansatz, we get three differential
equations. Those for $Z\left(z\right)$ and $\Phi\left(\varphi\right)$ are:
\begin{equation}
  \frac{d^{2}Z\left( z\right) }{dz^{2}}+k_{z}^{2}Z\left( z\right)
  =0,
\end{equation}
\begin{equation}
  \frac{d^{2}\Phi\left( \varphi\right)
  }{d\varphi^{2}}+m^{2}\Phi\left(
    \varphi\right)  =0,
\end{equation}
whose solutions, in complex form, are respectively
\begin{equation}
  Z\left(  z\right)  =e^{\pm ik_{z}z},%
\end{equation}
and %
\begin{equation}
  \Phi\left(  \varphi\right)  =e^{\pm im\varphi}.
\end{equation}
The third differential equation,
corresponding to the radial part $R\left(\rho\right) $ of the ansatz,
is the Bessel differential equation:

\begin{equation}
  \frac{d^{2}}{d\rho^{2}}R\left( \rho\right)
  +\frac{1}{\rho}\frac{d}{d\rho
  }R\left(  \rho\right)  +\left[ \left(  k^{2}-k_{z}^{2}\right)
    -\frac{m^{2}}{\rho^2}\right]  R\left(  \rho\right)  =0,
  \label{ecdfb}%
\end{equation}
where we can define $k_{\rho}=k^{2}- k_{z}^{2}$ and the solutions in
complex form are given by
\begin{equation}
  H_{m}^{\left( 1\right) }\left( k_{\rho}\rho\right) =J_{m}\left(
    k_{\rho
    }\rho\right)  +iN_{m}\left(  k_{\rho}\rho\right),
  \label{h1}%
\end{equation}
and
\begin{equation}
  H_{m}^{\left( 2\right) }\left( k_{\rho}\rho\right) =J_{m}\left(
    k_{\rho
    }\rho\right)  -iN_{m}\left(  k_{\rho}\rho\right),  
  \label{h2}%
\end{equation}
which are known as the Hankel functions of the first and second kind,
respectively. The Hankel functions are singular due to the presence of
the singularity of the Neumann function. However we will show below
that this singularity has an actual physical meaning. Notice that
Equation (\ref{h1}), represents the Green's function of the Helmholtz
equation in cylindrical coordinates: it implies that we have a source
of light that emanates energy radially.

\begin{figure}[h!]
  \centering
  \includegraphics[width=8.6 cm]{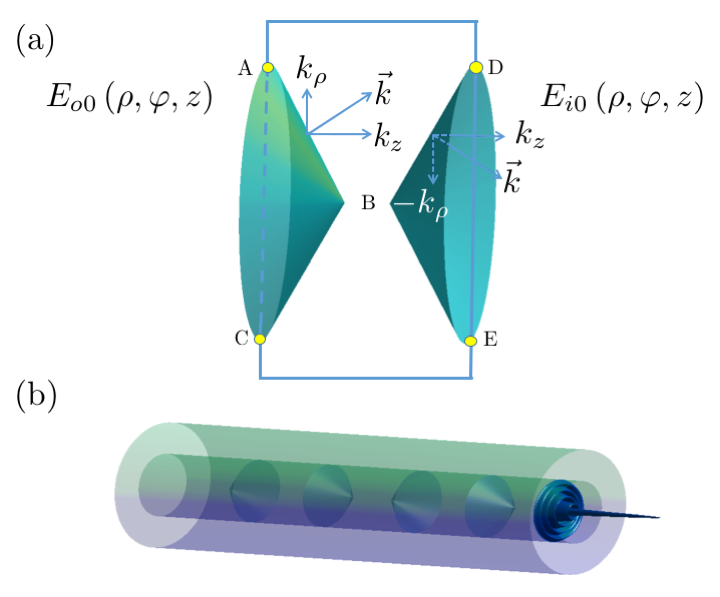}
  \caption{a) Representation of conical counter-propagating waves
    which are solutions to the Helmholtz equation in a cylindrical
    waveguide. They generate Hankel functions. b) The sum of
    conical counter-propagating waves generate a Bessel function of
    order zero.}
  \label{fig:Cyl}
\end{figure}

We can now focus our attention on the physics of these solutions. To simplify the description and the visualization let
  us take $m=0$; the observations below still apply for any
  $m$. Incorporating the $z-$dependence we have
  $H_{0}^{\left(1\right)}\left(k_{\rho}\rho\right)e^{ikz}$ and
  $H_{0}^{\left(2\right)}\left(k_{\rho}\rho\right)e^{ikz}$. These
  equations represent conical waves with a total wavevector
  $\overrightarrow{k}=k_{\rho}\widehat{\rho}+k_{z}\widehat{z}$, see
  Fig. \ref{fig:Cyl} a). The zero-th order Hankel function of the first kind
  $H_{0}^{\left(1\right)}$, describes radially symmetric cylindrical
  waves traveling away from the axis (outgoing waves). The radial
  component of the outgoing conical wave on reflection at the surface
  of the waveguide becomes an incoming conical wave described by
  $H_{0}^{\left(2\right)}$, i.e., the zero-th order Hankel function of
  the second kind $H_{0}^{\left(2\right)}$, represents waves traveling
  towards the axis (incoming waves). This is easy to visualize if the propagation of the conical wavefronts are known. The evolution of the propagation of a transverse section of the conical wavefronts are shown in Fig. 2, where the red dots represent outgoing conical wave while the black dots incoming conical wave.
 It is possible to observe that a reflected incoming conical wave is generated when the outgoing one has reached the border and that the incoming conical wave is transformed into an outgoing wave when the former passes through the axis of the waveguide. It is important to say that the transverse section of the conical wavefronts is generated by the points ACDE from Fig. 1, and the transverse section of the wavefronts is divided in representative sections, solid circles in Fig. 2.

\begin{figure}[h!]
  \centering
  \includegraphics[width=8.6 cm]{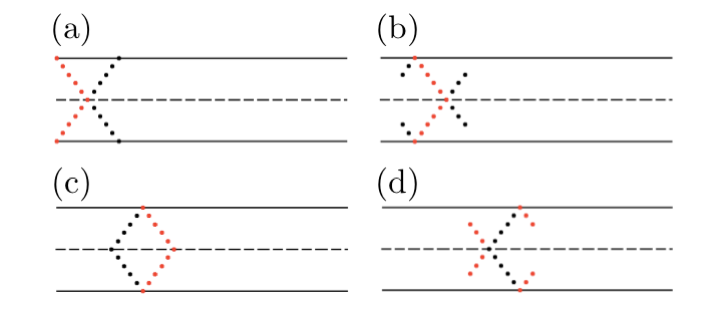}
  \caption{The propagation evolution of the conical wavefronts.} 
  \label{fig:Cosine_Bessel}
\end{figure}

Since each wave is the complex conjugate of the
  other, the incoming and outgoing waves have the same frequency and amplitude, and consequently, their transverse radial components move in
opposite directions, and in superposition the singular Neumann functions cancel
  out. An alternative way to explain this fact is by means of
  mathematical physics: the cylindrical waveguide supports cylindrical
  incoming waves traveling towards the longitudinal axis
  ($\rho=0$). As they get closer and closer to the axis, the waves
  ``collapse'' into a line, and it is actually this line which
  simultaneously acts as the source from which the outgoing
  cylindrical waves emanate. This is the physical explanation of the
  singularity of the Hankel functions, within the waveguide there are simultaneously a sink
  and a source that cancel out resulting in the $J_0$ non-singular stationary wave solution
  \cite{chavez}.  In the core of the cylindrical waveguide both waves exist simultaneously, this is, the solution must be given by
  
\begin{equation}
E\left(\rho,z\right)=\left[H_0^{\left(1\right)}\left(k_{\rho}\rho\right)
   +H_0^{\left(2\right)}\left(k_{\rho}\rho\right)\right]e^{ik_z z}
   =2J_{0}\left(k_{\rho}\rho\right)e^{ik_z z}.
  \label{solH}
\end{equation}

In order to expose the propagating wave behavior described by the
Hankel functions we introduce its asymptotic approximation and the harmonic temporal dependence $\exp(-i\omega t)$ of the wave equation is reincorporated \cite{Arfken}:
\begin{equation}
  H_{m}^{\left(1,2\right)}\left(k_{\rho}\rho\right)e^{ik_z z}\approx
  \frac{A}{\sqrt{k_{\rho}\rho}}
  e^{-i(\omega t \mp k_{\rho}\rho+ik_z z)-i\frac{\pi}{2}\left(m+\frac{1}{2}\right)}.
\end{equation}
From this expression it is clear the traveling wave behavior and the conical nature of the wavefronts.

\begin{figure}[h!]
  \centering
  \includegraphics[width=8.6 cm]{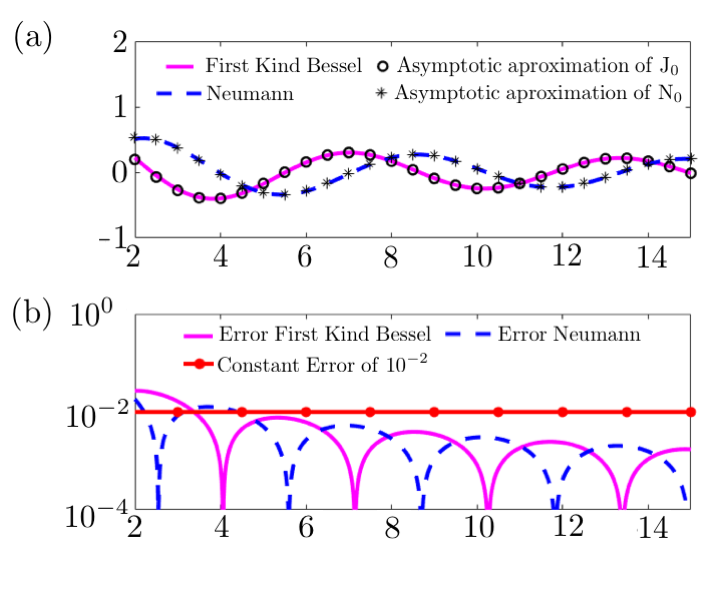}
  \caption{a) The magenta continuous line is the First Kind Bessel functions and the blue dashed-line is the Neumann functions. The black asterisks and the black circles represent the asymptotic approximation of the First Kind Bessel functions and the asymptotic approximation of the Neumann functions, respectively. b) The red horizontal line represents an error of $10^{-2}$. The magenta continuous line represents the error difference between the asymptotic approximation and the First Kind Bessel functions. The blue dashed-line represents the error difference between the asymptotic approximation of the Neumann functions and the Neumann functions. }
  \label{fig:Asym_Aprox}
\end{figure}

One may wonder how big has to be the argument of the Hankel functions to the asymptotic expression to be valid. One finds in the literature (see e.g. Ref.[17]) that this expression can be used when $k_{\rho}\rho\gg\left(4m^{2}-1\right) /8$ with $m\geq1$. However, giving a quantitative value, if we require that the error be of the order of $10^{-2}$ or less we have to increase $k_{\rho}\rho$ by at least a factor of five:%
\begin{equation}
  k_{\rho}\rho\geq 5\frac{\left|  4m^{2}-1\right|  }{8},%
\end{equation}
this approximation is now valid for representing the Bessel functions of
the first kind as well as for the Neumann functions, including the zero order ones. This shown in Fig. \ref{fig:Asym_Aprox} for $m=0$ where in the top plot the Bessel and Neumann functions together with their asymptotic are superposed and the simple difference is shown in in bottom plot.%

In the slab waveguides there exist incident and reflected waves inside it, as described in Refs. [4] and [5]. In analogy to the latter, we notice that inside the cylindrical waveguide there also exist incident as well as a reflected waves at and from the cylindrical wall of the waveguide. In both cases, when the waves are added together they form stationary waves inside the waveguide that are discussed in the literature for slab waveguides and we introduce for the first time for cylindrical waveguides. From equation (\ref{solH}), the solution can be
defined as the sum of conical counter-propagating waves.

It is well accepted that the stationary modes inside a slab waveguide are generated by the sum of counter-propagating waves bouncing back and forth from the plane walls of the slab. With this in mind we introduce another approach to describe the stationary waves inside a cylindrical waveguide from the picture of the modes of the slab waveguide. The latter is possible by rotating the slab waveguide $\pi$ radians, as it is shown in Ref. [18]. Since a plane wave in a polar coordinates can be represent by $e^{ik_{\rho}\left( x\cos\varphi+y\sin\varphi\right)+i k_z z}$ the continuous interference of these counter-propagating plane waves in the $\pi$ rotation can be represent by

\begin{equation}
  \frac{1}{\pi}\int_{0}^{\pi}\cos{k_{\rho }
  \left(  x\cos\varphi+y\sin\varphi\right)}e^{ i k_z z}d\varphi =
  \frac{1}{2\pi}\int_{0}^{2\pi}e^{ik_{\rho }
  \left(  x\cos\varphi+y\sin\varphi +i k_z z\right)}d\varphi 
  \equiv J_{0}\left( k_{\rho}\rho\right) e^{ik_z z}.
  \label{intJ}
\end{equation}

In a similar way, under a change of variables it can be demonstrated that if the plane waves have an azimuthal phase shift of the form $e^{im\varphi}$ as they rotate the resulting field is described by

\begin{equation}
  J_{m}\left( k_{\rho}\rho\right) e^{-i m \theta +i k_z z}
  =\frac{1}{2\pi}\int_{0}^{2\pi}e^{ik_{\rho}\rho
  \left(  \cos\alpha+\sin\alpha\right ) - im\alpha +i k_z z }d\alpha.
\end{equation}
where $\theta$ is now the azimuthal variable.

For $m=0$, the wavefront generated by the rotation of the plane waves forms two conical surfaces; cone A, B, C represents the outgoing conical wave incident to the inside of the waveguide, and cone D, B, E represents the
reflected conical incoming wavefront. This is easy to visualize due to the
change in sign of the $k_{\rho}$ component, see Fig. \ref{fig:Cyl}.

From Fig. \ref{fig:Cyl} we can see that the vector
$\overrightarrow{k}$ and its components $k_{\rho}$ and $k_{z}$ are
located in the sagital plane of the cylinder,
i.e. the vectors are coplanar vectors. Notice that the integral in \ref{intJ} 
creates a the cone of wavevectors. 

Also observe that using the rotation of the planar 
slab description, the singular Neumann function 
cannot be constructed and this solution is discarded in a 
natural way. Physically, we can also deduce this from equations (\ref{h1}) and (\ref{h2}) since in order to get the Neumann solution we would require to subtract the waves, what would imply that there was a relative phase of $\pi$ between the incomimg and outgoing conical waves along the whole waveguide, something that does no occur. In this manner the approach presented here contributes to making the modes inside the cylindrical waveguide more physically understandable.

Now, we will demonstrate how the traveling waves
  inside the core become, in a natural way, the evanescent wave at the
  cladding.  We have said that in the core
  $H_0^{\left(1\right)}\left(k_{\rho}\rho\right)e^{ikz}$ represents
  the outgoing wave. At the cylindrical waveguide surface $\rho=a$ the
  incident wave, by total reflection, must give rise to an evanescent field
  outside the core. In the same way as occurs with the slab waveguide,
  the transverse wave number becomes purely imaginary, i.e.
  $k_{\rho t}= i\kappa_{\rho t}$, resulting in having the Hankel
function of the first type to be transmitted to the outer part of the
cylindrical waveguide. In this case the solution in this region is
given by
\begin{equation}
  E\left(\rho>a,z\right)=H_{0}^{\left( 1\right)
  }\left(  i\kappa_{\rho}\rho\right)e^{ik_{z}z}.
  \label{SolOutOut}
\end{equation}
This equation can be rewritten using the modified Bessel function of
the second kind, i.e.,
$K_{m}\left( \kappa_{\rho}\rho\right)
=\frac{\pi}{2}i^{m+1}H_{m}^{\left( 1\right)
}\left(i\kappa_{\rho}\rho\right)$
\cite{Arfken,Boas}, and for $m=0$ we have
\begin{equation}
  E\left(\rho>a,z\right) = {\frac{2}{\pi}K_{0}\left(\kappa_{\rho}\rho\right)}e^{i\left(
        k_{z}z+\frac{\pi}{2}\right)}.
\label{CylindricalCladding}
\end{equation}
This result demonstrates how the outgoing traveling wave transforms in
a natural way in an evanescent wave described by the modified Bessel
function $K_0(\kappa_{\rho} \rho)$.

Now, the transmitted wave just at the cladding is 
\begin{equation}
J_{0}\left(\kappa_{\rho}a\right)={\left|\frac{e^{i\frac{\pi}{2}}}{\pi} K_{0}\left(\kappa_{\rho}a\right)\right|},
\end{equation}
where we can observe that the amplitude coefficient of the transmitted
wave is given by $e^{i\frac{\pi}{2}}/\pi$. Moreover, this $\pi/2$
phase shift can be interpreted as the rotation of the radial component
of the wavevector to create the surface waves.  

We remark that it was not necessary to make any mathematical
assumptions in order to end up with evanescent waves with
Eq. (\ref{CylindricalCladding}) at the surface of the cylindrical
waveguide. Contrary to what is done in the physical mathematics approach
where it is usually that the modified Bessel function $K_m$ is chosen
over the modified Bessel function $I_m$ because the latter grows to infinity 
while the former one decays. In our analysis the direct use of the Hankel
function as a solution inside the waveguide gives rise naturally to
the evanescent wave in the form of the modified Bessel function $K_m$
without having to make any further mathematical assumptions. Furthermore, in
the core, the solutions we have obtained are also able to describe the
phenomena observed in tubular mirrors \cite{Cohen, Yves}.

As it has been seen throughout this section, for sake of clarity the order of the Hankel function has been taken as zero, i.e., $m= 0$. In the following section we will be briefly describe the solution of the wavefronts of higher order modes $m\neq0$ inside of the cylindrical waveguide. These modes are also known as skew modes in the Optical Fibers literature.

\section{Higher-order modes in cylindrical waveguides}
We will now turn our attention to the higher-order modes in
cylindrical waveguides, which have the solution: a) in the core
\begin{equation}
  \begin{array}
    [c]{c}
    E_{om}\left(\rho,\varphi,z\right) =
    e^{-i\left(k_{z}z+m\varphi\right)}H_{m}^{\left(1\right)}\left(k_{\rho}\rho\right)\\
    E_{im}\left(\rho,\varphi,z\right) =
    e^{-i\left(k_{z}z+m\varphi\right)}H_{m}^{\left(2\right)}\left(k_{\rho}\rho\right),
  \end{array}
\end{equation}
The factor $e^{-im\varphi}$ represents that the phase rotates $m$ times in a period of 0 a $2\pi$, as is shown in Fig. 4.

b) In the cladding
\begin{equation}
  E\left(\rho,\varphi,z\right)={\frac{2
    }{\pi}e^{-i\left(
        k_{z}z+m\left[\varphi-\frac{\pi}{2}\right]-\frac{\pi}{2}\right)
    }K_{m}\left(
      \kappa_{\rho}\rho\right)}.
\end{equation}

\begin{figure}[h!]
  \centering
  \includegraphics[width=8.6 cm]{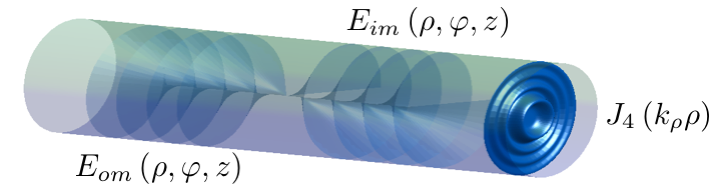}
  \caption{High-order mode in cylindrical waveguides with $m=4$ and its conical helicoidal wavefronts.}
 \label{fig:High_order}
\end{figure}

We have explained that lower order modes in a cylindrical waveguide
have a conical wavefront. For the case of higher order modes, the
wavefront is a conical helicoid, as show in Fig. \ref{fig:High_order}.

As the wavefront is a conical helicoid, the vector of
propagation$\overrightarrow{k}$ follows a screw-like helical trajectory and its components no longer lie on a plane perpendicular to the tangential plane of the cylinder at any given point.

\section{Conclusions}

We have presented a discussion of the differences between the physical
mathematics and mathematical physics approaches used to describe
problems in physics. We have demonstrated that using the mathematical
physics approach some physical aspects of the propagation of
electromagnetic waves in cylindrical waveguides are recovered that
other wise using the physical mathematics would be lost.

Having the traveling wave idea in mind, we demonstrated that the waves
inside the cylindrical waveguide are in fact propagating conical
waves.  From a physical mathematics point of view these conical waves
would never be seen, since they are described with the Hankel
functions of the first and second kinds, that have a singularity due
to the presence of the singular Neumann functions.  Nonetheless, with
the aid of the mathematical physics picture, we have shown that the
Neumann functions are a very important component in the full
description of the physics of these conical waves. The standing waves
inside the cylindrical waveguide, which have profile given by the
Bessel function of the first kind, is the result of the transverse
counter-propagating component of the conical waves described with the
Hankel functions. The counter-propagation results in the superposition
of the incoming and outgoing waves canceling out the term with the
Neumann functions in a natural manner, without the need of arbitrarily
discarding them. Also, we demonstrated how the outgoing wave
transforms in a simple straightforward manner in the evanescent field
when the condition for total internal reflection is satisfied.

In general, the physics oriented method presented in this paper, the
mathematical physics approach, gives a more physically rich insight of
the modes inside cylindrical
waveguide than more prevalent physical mathematics methods in the
literature.

\begin{acknowledgments}

The first author would like to acknowledge J. A.
Carbajal-Dom\'inguez and R. J. Le\'{o}n-Montiel for fruitful discussions.
\end{acknowledgments}

\end{document}